\documentclass[11pt]{article}
\pdfoutput=1

 \usepackage{jcappub}
\usepackage{caption}
\usepackage{subcaption}
\usepackage{hyperref}
\usepackage{graphicx}
\usepackage{amsmath}
\usepackage{mathtools}
\usepackage{physics}
\usepackage{siunitx}
\usepackage{tablefootnote}
\usepackage{soul}
\usepackage{color}
\usepackage{verbatim}
\usepackage{float}
\usepackage{bm}
\usepackage[normalem]{ulem}
\usepackage[dvipsnames]{xcolor}
\usepackage{tikz} 
\usepackage{tikz-feynman}
\usepackage{orcidlink}
\usepackage[nameinlink, capitalise]{cleveref}
\usepackage{braket}
\usepackage{dirtytalk}
\usepackage{mathtools}
\usepackage{amsmath}
\usepackage{tikz}
\usepackage{tikz-feynman}
\tikzfeynmanset{compat=1.1.0}
\usepackage{feynmp}
\usetikzlibrary{shapes.geometric, backgrounds, arrows, calc, fit}
\tikzstyle{process} = [rectangle, rounded corners, minimum width=8cm, minimum height=1cm, text centered, draw=black, fill=blue!30]
\tikzstyle{arrow} = [thick,->,>=stealth]
\usepackage{underscore}

\usepackage{color,appendix,latexsym,amsmath,amssymb,graphicx,booktabs,epsfig,hyperref,url}

\numberwithin{equation}{section}
\usepackage[english]{babel}

\usepackage{letltxmacro}
\LetLtxMacro{\originaleqref}{\eqref}
\usepackage[sorting=none]{biblatex} 
\addto\extrasenglish{%
}

\newcommand{\kt}{\tilde{k}}

\definecolor{MyBlue}{rgb}{0.15,0.15,0.70}
\hypersetup{
colorlinks=true,
citecolor=MyBlue,
linkcolor=MyBlue,
urlcolor=MyBlue
}

\definecolor{orange}{rgb}{0.98, 0.6, 0.01}
\definecolor{darkolivegreen}{rgb}{0.33, 0.42, 0.18}
\definecolor{tealblue}{rgb}{0.21, 0.46, 0.53}

\usepackage{xspace}

\newcommand{\fpbh}{\ensuremath{f_{\rm PBH}}}
\newcommand{\fnl}{\ensuremath{\tilde{f}_{\rm NL}}}
\newcommand{\gnl}{\ensuremath{\tilde{g}_{\rm NL}}}
\newcommand{\betaz}{\ensuremath{\beta_{1-\rm PBH}}}
\newcommand{\calR}{\ensuremath{{\cal R}}}

\newcommand{\erfc}{\ensuremath{{\rm ErfC}}}
\newcommand{\inverfc}{\ensuremath{{\rm InvErfC}}}

\usepackage{listings}
\definecolor{codegreen}{rgb}{0,0.6,0}
\definecolor{codegray}{rgb}{0.5,0.5,0.5}
\definecolor{codepurple}{rgb}{0.58,0,0.82}
\definecolor{backcolour}{rgb}{0.95,0.95,0.92}
\graphicspath{figures/} 

\title{Robust $\mu$-distortion constraints on primordial supermassive black holes from cubic ($g_{NL}$) non--Gaussian perturbations}

\author[1]{Xavier Pritchard\,\orcidlink{0009-0007-6543-8563},}
\author[1]{Christian T.~Byrnes\,\orcidlink{0000-0003-2583-6536},}
\author[2]{Julien Lesgourgues \,\orcidlink{0000-0001-7627-353X},}
\author[2]{Devanshu Sharma\,\orcidlink{0009-0002-3302-2153}}

\affiliation[1]{ Department of Physics and Astronomy, University of Sussex, Brighton BN1 9QH, UK\\}
\affiliation[2]{Institute for Theoretical Particle Physics and Cosmology (TTK), RWTH Aachen University, \\ D-52056 Aachen, Germany}

\emailAdd{X.Pritchard@sussex.ac.uk}
\emailAdd{C.Byrnes@sussex.ac.uk}
\emailAdd{lesgourg@physik.rwth-aachen.de}
\emailAdd{drsharma@physik.rwth-aachen.de}

\date{}

\abstract{We make the first calculation of the spectral distortion constraints on the primordial curvature power spectrum in the limit of large cubic non--Gaussianity. This calculation involves computing a 2-loop integral, which we perform analytically. Despite being non-perturbatively non--Gaussian, we show that the constraints only change significantly from the case of Gaussian perturbations in the high--$k$ tail, where spectral distortions become weak. We conclude that generating primordial supermassive black holes requires even more extreme forms of non--Gaussianity. We also argue why the $\mu$-distortion constraint is unlikely to significantly change even in the presence of more extreme local non--Gaussianity.}

\begin{document}

\begin{flushleft}
TTK-25-13
\end{flushleft}

\maketitle
\flushbottom

\section{Introduction}

The study of distortions in the frequency spectrum of the Cosmic Microwave Background (CMB) offers a fascinating view of the early and late universe, which is complementary to the information encoded in the CMB spatial temperature anisotropies \cite{Sunyaev:1970er,Chluba:2011hw,1975SvA....18..413I}. The observations carried out by the COBE FIRAS instrument found no deviations in the blackbody spectrum of the CMB up to $\mathcal{O}(10^{-5})$ \cite{Mather:1990tfx,Mather:1993ij,Fixsen:1996nj}. However, the presence of distortions at lower magnitudes could provide crucial information about a number of cosmological and astrophysical events, including the possible detection of primordial magnetic fields, broader constraints on the gravitational wave background from scalar induced gravitational waves and deeper insights into 21cm cosmology \cite{Kogut:2019vqh,Cyr:2023pgw,Tagliazucchi:2023dai,Cyr:2024pme,Li:2025met}. Consequently, several missions have been proposed that are expected to probe the spectral distortions with precision as high as $\mathcal{O}(10^{-8})$ \cite{Chluba:2019nxa,Maffei:2021xur,Kogut:2024vbi}.

Spectral distortions are generated because of the reduced efficiency of the photon-baryon fluid to redistribute external energy evenly throughout all frequencies. It is well known that photon diffusion in the radiation-dominated era occurs due to the phenomenon of Silk damping \cite{Silk:1967kq,Jeong:2014gna}. Not only does it lead to the diffusion of primordial perturbations (stored as photon temperature fluctuations), but it also causes the blending of multiple photons at slightly different temperatures. The latter effect gives rise to distortions in the CMB because, in the $\mu$ era ($5 \times 10^4 < z < 2 \times 10^6 $), the production of photons through Bremsstrahlung becomes inefficient. As the universe cools further until $z \leq 5 \times 10^4$, even the photon-baryon interaction through Compton scattering loses efficiency and we have the $y$ era. For more on the theoretical and mathematical framework of the spectral distortions, the reader is requested to refer to \cite{Chluba:2012we,Chluba:2012gq,Pajer:2012qep,Khatri:2012rt,Tashiro:2014pga,Kohri:2014lza,Lucca:2019rxf,Sharma:2024img}.

Current observations do not tightly constrain the statistics of inflationary fluctuations on smaller scales $k>\mathcal{O}(0.05 \ {\rm Mpc}^{-1})$. Consequently, the primordial curvature power spectrum may be enhanced by many orders of magnitude, leading to large density contrasts between the fluctuations and plasma. Primordial black holes (PBHs) are then able to form via the gravitational collapse of extreme density fluctuations \cite{Zeldovich:1967lct,Hawking:1971ei,Carr:1974nx,Chapline:1975ojl}. Large enhancements in the fluctuation amplitude can be realised in a plethora of inflation models beyond single-field slow-roll, ranging from ultra-slow-roll models to string theory motivated models; see e.g.~\cite{Garcia-Bellido:1996mdl,Clesse:2015wea,Garcia-Bellido:2017mdw,Mishra:2019pzq,Ballesteros:2020qam,Gangopadhyay:2021kmf,Cole:2023wyx,Stamou:2024lqf,Yogesh:2025hll}.

The observation (or lack thereof) of spectral distortions, particularly $\mu$-distortions, caused by the dissipation of primordial scalar perturbations can therefore help constrain the abundance of PBHs at present times. Interestingly, electron-position annihilation, which may lead to a natural enhancement in PBH abundances via a softening in equation of state \cite{Jedamzik:1996mr,Carr:2019kxo,Musco:2023dak,Pritchard:2024vix}, also occurs within the range of scales probed by spectral distortions. Furthermore, the resulting PBH masses lie in the supermassive range. Thus, $\mu$-distortions serve as a crucial tool in understanding whether supermassive black holes are of primordial origin (see also 
\cite{Papanikolaou:2023cku,Papanikolaou:2023nkx}). 

As the density perturbations in the post-inflationary universe are passed on from the vacuum fluctuations of the scalar field driving inflation, one can assume that each mode of the density fluctuation in Fourier space is an independent harmonic oscillator that follows a Gaussian probability distribution. This notion, however, is not entirely correct if one accounts for the large fluctuations that lead to PBH formation. Such fluctuations tend to interact at different scales and lose the status of independent harmonic oscillators and thus, the assumption that they follow a Gaussian distribution does not stay valid \cite{Byrnes:2014pja,Meerburg:2019qqi,Gow:2022jfb,Ferrante:2022mui}.

In a series of two past papers, we made the first calculation of the spectral distortion constraints on the power spectrum amplitude without assuming Gaussian perturbations \cite{Sharma:2024img, Byrnes:2024vjt}. We extended the existing constraint for Gaussian perturbations to local-type quadratic non--Gaussianity and studied both the perturbative and non-perturbative limits. In doing so, we showed that the tightest $\mu$-constraint barely changes from its Gaussian value even in the limit of a pure $\chi^2$ non--Gaussianity and provided an intuitive reason why this picture might not change much even in the case of stronger non--Gaussianities. As expected, changes to the constraint for perturbative non--Gaussianity were even smaller than in the non-perturbative case. However, there are significant changes towards the tails of the constraints where there is more sensitivity to the shape of the effective power spectrum, which varies between the Gaussian and non--Gaussian cases. 

In this work, we extend our previous results by determining the power spectrum constraints for cubic local non--Gaussianity, focusing on the infinite $g_{NL}$ limit. We show that even for this extreme form of non--Gaussianity, the constraints remain remarkably similar to the Gaussian case. We provide an intuitive reason and strengthen our previous arguments about why this should remain true to all higher-orders in non--Gaussianity. We therefore pave the way to making robust constraints on the formation of primordial supermassive black holes (SMBHs) since the power spectrum amplitude required to generate PBHs can be determined for the relevant extreme forms of local non--Gaussianity to all orders. 

A work by other authors has shown that our previous constraints, which were derived assuming a Dirac delta function for the Gaussian power spectrum remained robust for broader peaks, and the qualitative picture did not change whether they used a lognormal or broken power law peak \cite{Wang:2025hbw}  (see the journal version) (also in agreement with \cite{Yang:2024snb} who used a compaction function analysis). This remained true for Gaussian and non--Gaussian perturbations, and we expect this to remain true even with higher-order non--Gaussianity.

The rest of the paper is organised as follows: In \cref{cubicnonGSection} we calculate the 2-loop non--Gaussian power spectrum, corresponding to cubic non--Gaussianity. In \cref{ConstraintsSection} we discuss the corresponding $\mu$-distortion constraints, as well as relevant constraints on PBH abundances. We conclude in \cref{conclusion} and provide a detailed description of our calculation of the power spectrum in \cref{detailedSection}. In \cref{pureSunsetSection} we add some discussion about individual loop diagram contributions and in \cref{infiniteLoopSection} we discuss the extension of our work to arbitrary higher--order non--Gaussianity. 

\section{Cubic non--Gaussianity}\label{cubicnonGSection}

Local type primordial non--Gaussianity is often parametrised as a Taylor expansion of the comoving curvature perturbation,
\begin{equation}
    \mathcal{R}(\vec{x})=\mathcal{R}_\text{G}(\vec{x})+\tilde{f}_\text{NL}\left(\mathcal{R}_\text{G}(\vec{x})^2-\braket{\mathcal{R}_\text{G}(\vec{x})^2} \right)+\tilde{g}_\text{NL}\mathcal{R}_\text{G}(\vec{x})^3+\cdots, \label{R-expansion}
\end{equation}
where $\tilde{f}_\text{NL}\equiv3f_\text{NL}/5$ and $\tilde{g}_\text{NL}\equiv9g_\text{NL}/25$ are expansion coefficients and $\mathcal{R}_\text{G}$ is a Gaussian random field typically with $\abs{\mathcal{R}_\text{G}(\vec{x})}\ll\mathcal{O}(1)$ and $\braket{\mathcal{R}_\text{G}(\vec{x})}=0$. Recently, $\mu$-distortion constraints have been studied in the context of non--Gaussianity arising from $\tilde{f}_\text{NL}$ \cite{Sharma:2024img, Byrnes:2024vjt, Hai-LongHuang:2024vvz}. Regarding \cref{R-expansion}, the pure $\chi^2$ contribution coming from the infinite $\fnl$ limit is equivalent to setting each expansion coefficient to zero, except for the $\tilde{f}_\text{NL}$ term. Using a similar approach, we investigate $\mu$ constraints in the context of pure cubic non--Gaussianity. In this scenario, we only need to consider the $\tilde{g}_\text{NL}$ contribution.

In Fourier space, this corresponds to setting the curvature perturbation as follows
\begin{equation} 
     \mathcal{R}(\vec{k})=\tilde{g}_\text{NL}\int\frac{d^3\vec{q_1}}{(2\pi)^{3/2}}\frac{d^3\vec{q_2}}{(2\pi)^{3/2}}\mathcal{R}_G(\vec{q_1})\mathcal{R}_G(\vec{q_2})\mathcal{R}_G(\vec{k}-\vec{q_1}-\vec{q_2}),
\end{equation}
where the Gaussian component has been omitted, which we justify in later discussion. This corresponds to non-perturbative non--Gaussianity, in which the curvature perturbation is dominated by the non--Gaussian $\tilde{g}_\text{NL}$ term. In the limit that this term is infinite, a pure Gaussian cubed distribution is recovered.

\begin{figure}[ht!]
  \begin{minipage}[b]{0.4\textwidth}
   \begin{tikzpicture}[scale=3, transform shape]
      \begin{feynman}
        \vertex (a); 
        \vertex [right = 0.55cm  of a] (b);
        \vertex [below = 0.506cm of a] (a1);
        \vertex [right = 0.8cm  of b] (c);
        \vertex [right = 0.55cm  of c] (d);
        \vertex [above = 0.5cm  of b] (e);
        \vertex [above = 0.5cm  of c] (f);

        \diagram* {(a) -- [] (b), 
          (a) -- [opacity=0] (a1),
          (b) -- [scalar, half right, looseness=1.5] (e), 
          (b) -- [scalar, half left, looseness=1.5] (e), 
          (b) -- [scalar, looseness=0] (c),
          (c) -- [scalar, half right, looseness=1.5] (f), 
          (c) -- [scalar, half left, looseness=1.5] (f), 
          (c) -- [] (d),
          };
      \end{feynman}
\end{tikzpicture}
  \end{minipage}
  \hspace{2mm} 
  \begin{minipage}[b]{0.4\textwidth}
   \begin{tikzpicture}[scale=3, transform shape]
       \begin{feynman}
        \vertex (a); 
        \vertex [right = 0.55cm  of a] (b);
        \vertex [right = 1.3cm  of b] (c);
        \vertex [right = 0.55cm  of c] (d);

        \diagram* {
          (a) -- [] (b), 
          (b) -- [scalar, half right, looseness=1] (c), 
          (b) -- [scalar, half left, looseness=1] (c), 
          (b) -- [scalar] (c),
          (c) -- [] (d)
          };
      \end{feynman}
\end{tikzpicture}
  \end{minipage}
  \caption{The 2-loop terms in our calculation. {\it Left}: The simpler term, involving no convolution integrals. We refer to this diagram as the dressed vertex diagram. {\it Right}: The more complicated 2-loop integral, the calculation of which is one of the main results of the paper. We refer to this diagram as the sunset diagram.}\label{DiagramPlot1}
\end{figure}
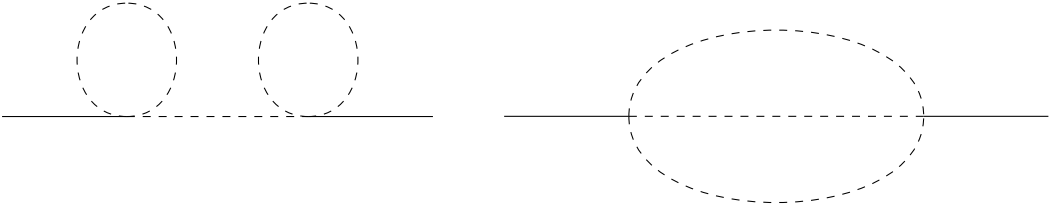

Focusing on the primordial power spectrum, we are able to compute the leading non--Gaussian correction to spectral distortions. Specifically, we assume that the Gaussian fluctuations can be described by a power spectrum parametrised by the Dirac-delta function
\begin{equation}
    \mathcal{P}_G(k)\equiv\frac{k^3}{2\pi^2}P_G(k)=A_Gk_*\delta(k-k_*),
\end{equation}
where $A_G$ is the power spectrum amplitude and $k_*$ defines the peak scale. In \cite{Wang:2025hbw}, constraints arising from broader power spectra were considered; however, this makes the calculation more involved and did not change the main conclusions. With this assumed shape for the Gaussian power spectrum and following \cite{Byrnes:2007tm}, we are able to describe the non--Gaussian power spectrum at 2-loop order. This is given by computing a pair of diagrams, as shown in \cref{DiagramPlot1}.

\subsection{The dressed vertex diagram}

The dressed vertex diagram corresponds to \cite{Byrnes:2007tm} 
$$ P_{G^3,DV} = 6^2\frac14  \int \frac{d^3q_1}{(2\pi)^3}\frac{d^3q_2}{(2\pi)^3} P_G(q_1)P_G(q_2)P_G(k) $$
where the numerical factor of $6^2$ comes from the fact that there is no $1/3!$ prefactor to the third--order term in \cref{R-expansion}. The integration is straightforward due to there being no convolution integrals.
Noting that
\begin{equation}
    \int \frac{d^3q}{(2\pi)^3} P_G(q)=4\pi \int \frac{dq}{(2\pi)^3}\frac{2\pi^2}{q^3}A_G k_*\delta(k-k_*)=A_G
\end{equation}
we see that this 2-loop diagram reduces to
\begin{eqnarray}
 {\cal P}_{G^3,DV}=9A_G^3 k_* \delta(k-k_*)
\end{eqnarray}
so it has the same shape as the Gaussian power spectrum but its amplitude is suppressed by a factor of $9A_G^2$. 

\subsection{The sunset diagram}\label{sunsetsection}

Here we sketch the main steps in the derivation of the analytic solution to the complicated 2-loop integral. Further details are in \cref{detailedSection}. 
The sunset diagram corresponds to \cite{Byrnes:2007tm}
\begin{equation}\label{eq:sunset-convolution}
P_{G^3,S}=6\int \frac{d^3q_1}{(2\pi)^3}\int \frac{d^3q_2}{(2\pi)^3}P_G(\vec q_1)P_G(\vec q_2)P_G(\lvert\vec k-\vec q_1-\vec q_2\rvert).
\end{equation}
Using spherical polar coordinates, this can be written as
\begin{align}\label{eq:sunset-q-magnitude}
        P_{G^3,S}=\frac{6}{(2\pi)^6}&\int^{2\pi}_0\text{d}\phi_1\int^{2\pi}_0\text{d}\phi_2\int^{\pi}_0\text{sin}(\theta_1)\text{d}\theta_1\int^{\pi}_0\text{sin}(\theta_2)\text{d}\theta_2\int^{\infty}_0q_1^2\text{d}q_1\int^{\infty}_0q_2^2\text{d}q_2\\
        & \left( \frac{2\pi^2A_Gk_*}{q_1^3}\delta(q_1-k_*)\right) \left(  \frac{2\pi^2A_Gk_*}{q_2^3}\delta(q_2-k_*)\right) \left(  \frac{2\pi^2A_Gk_*}{f^3}\delta(f-k_*)\right),\notag
\end{align}
where $f=\lvert\vec k-\vec q_1-\vec q_2\rvert$.

We are able to straightforwardly integrate over $q_1$, $q_2$ and $\phi_2$ as these variables are degenerate. Then, we change variables from $\theta_{1/2}$ to $x/y$ via the coordinate transformation $x/y$=cos$(\theta_{1/2})$, such that d$x/y$=sin($\theta_{1/2}$)d$\theta_{1/2}$. Collecting constant terms, the integral becomes
\begin{equation}
    P_{G^3,S}=\frac{3\pi A_G^3k_*}{2}\int^{1}_{-1}\text{d}x\int^{1}_{-1}\text{d}y\int^{2\pi}_0\text{d}\phi_1\frac{\delta(f-k_*)}{f^3}. \label{pre-phi-integration}
\end{equation}
Next, we define dot products as
\begin{equation}
    \vec q_1 \cdot \vec k = kq_1\text{cos}(\theta_1)\qquad   \vec q_2 \cdot \vec k = kq_2\text{cos}(\theta_2)\qquad   \vec q_1 \cdot \vec q_2 = q_1q_2\text{cos}(\theta_3),
\end{equation} 
where (see \cref{sec:theta3})
\begin{equation}
    \text{cos}(\theta_3)=\text{sin}(\theta_1)\text{sin}(\theta_2)\text{cos}(\phi_1)+\text{cos}(\theta_1)\text{cos}(\theta_2).
\end{equation}
This allows us to write the function within the Dirac-delta in the form
\begin{equation}\label{f-kstar}
    f-k_*=k_*\bigg(\sqrt{\kt^2-2\kt(x+y)+2(1+\sqrt{(1-x^2)(1-y^2)}\text{cos}(\phi_1)+xy)}-1\bigg),
\end{equation}
where we have introduced the dimensionless ratio, $\kt$, defined as
\begin{equation}
    \kt\equiv \frac{k}{k_*}.
\end{equation}
Focusing on the $\phi_1$ integral, we recall the definition of the $\delta$-function
\begin{equation}\label{DiracDelta}
\delta(F(\phi_1))=\Sigma_j\frac{\delta(\phi_1-\phi_j)}{|F'(\phi_j)|},\qquad \text{for}\quad F(\phi_j)=0,\quad F'(\phi_j)\neq 0.
\end{equation}
We show in \cref{sec:arcos} that
\begin{equation}
    F'(\phi_j)=\frac{k_*}{2}\sqrt{\alpha y^2 +\beta y +\gamma},
\end{equation}
with coefficients each being functions of $x$. We further impose the condition that the solutions to $F(\phi_j)=0$ are real, which is equivalent to
\begin{equation}
    \alpha y^2+\beta y +\gamma\geq0.\label{QuadraticInequality}
\end{equation}
Then, integrating with respect to $\phi_1$ leaves the following double integral
\begin{equation}
 \frac{k_*^3}{6\pi A_G^3}P_{G^3,S}=\int^{x_+(\kt)}_{x_-(\kt)}\text{d}x\int^{y_+(\kt)}_{y_-(\kt)}\text{d}y\frac{1}{\sqrt{ \alpha y^2+\beta y +\gamma}},
\end{equation}
where $x_\pm(\kt)$ and $y_\pm(\kt)$ are given as the values which satisfy \cref{QuadraticInequality}.
Focusing on the $y$-integral, we notice that it is of the form
\begin{equation}
   I = \int \frac{\text{d}y}{\sqrt{R}},
\end{equation}
where $R=\alpha y^2+\beta y+\gamma$. This integral has the following solutions \cite{Gradshteyn:1943cpj}
\begin{equation}
I=
 \begin{dcases} \label{arcsinsolution}
      \frac{1}{\sqrt{\alpha}}\text{ln}\left( \frac{2\sqrt{\alpha R}+2\alpha y+\beta}{\sqrt{\Delta}} \right)\quad\text{for}\quad \alpha>0, \\
      \frac{-1}{\sqrt{-\alpha}}\text{arcsin}\left( \frac{2\alpha y+\beta}{\sqrt{-\Delta}} \right)\quad\text{for}\quad \alpha<0,\quad\Delta<0, \\
   \end{dcases}
\end{equation}
where $\Delta=4ac-b^2$.

For the $x$ values considered, we find that $\alpha$ is always negative. Thus, if we are to obtain an analytic solution, we require $\Delta<0$. However, within our integration limits, this is the same requirement as \cref{QuadraticInequality}, which is discussed further in \cref{sec:arcos}. After plugging in the integration limits (see \cref{sec:arcsin}), we are left with the following integral
\begin{equation}
        \frac{k_*^3}{6\pi A_G^3}P_{G^3,S}=\frac{\pi}{2}\int^{x_+(\kt)}_{x_-(\kt)}\frac{\text{d}x}{\sqrt{1+\kt^2-2\kt x}},
\end{equation}
which we split into 2 cases. We note that these cases completely describe the power spectrum; $\kt<0$ is unphysical, whereas $\kt>3$ is prohibited due to momentum conservation. To be explicit, from \cref{eq:sunset-convolution} we require $\vec k+\vec q_1+\vec q_2=0$ and from \cref{eq:sunset-q-magnitude} the vectors we convolve over must have magnitude $q_1=q_2=k_*$.

\subsubsection{$0\leq\kt\leq1$}

For these values of $\kt$, we find that the integration limits, $x_\pm(\kt)$, are simply $\pm1$. In other words, the inequality \cref{QuadraticInequality} does not depend on the value of $x$ within the range of interest. The integral itself is simple to perform, leading to
\begin{equation}
    P_{G^3,S}=\frac{6A_G^3\pi^2}{k_*^3},
\end{equation}
such that
\begin{equation}
    \mathcal{P}_{G^3,S}=\frac{k^3}{2\pi^2}P_{G^3,S}=3\bigg( \frac{A_Gk}{k_*}\bigg)^3.
\end{equation}

\subsubsection{$1<\kt\leq3$}
Within this range of $\kt$, we find that the inequality \cref{QuadraticInequality}, is not satisfied for $-1<x<\frac{\kt^2-3}{2\kt}$. Thus, we set $x_-(\kt)=\frac{\kt^2-3}{2\kt}$ and $x_+(\kt)=1$. This leads to
\begin{equation}
    P_{G^3,S}=\frac{3A_G^3\pi^2}{\kt k_*^3}(3-\kt),
\end{equation}
which, finally, leads to
\begin{equation}
     \mathcal{P}_{G^3,S}=\frac{3A_G^3\kt^2}{2}(3-\kt)=\frac{3A_G^3k^2}{2k_*^3}(3k_*-k).
\end{equation}

\subsection{Variances of the power spectra}\label{VarianceSection}
%

Let us revisit the result of \cite{Sharma:2024img} obtained for the $\chi^2$ non--Gaussian contribution to the total comoving curvature perturbation $\mathcal{R} = \mathcal{R}_G + \tilde{f}_{\rm NL}\mathcal{R}^2_G$. For $\left|\tilde{f}_{\rm NL}\right| \gtrsim 1$, the leading order contribution to the spectral distortions in this case comes from the two-point correlator of the temperature anisotropy function i.e.~$\langle \Theta^2 \rangle$. With the help of the one-loop diagram \cite{Byrnes:2007tm}, the leading order correction to the power spectrum comes from the term which is proportional to $\mathcal{O}(\tilde{f}^2_{\rm NL})$
\begin{align}
\mathcal{P}(k) &= \mathcal{P}_{G}(k) + \tilde{f}^2_{\rm NL} \mathcal{P}_{G^2}(k)~, \\
&= \mathcal{P}_{G}(k) +      2 \frac{\tilde{f}_{\rm NL}^2}{(2\pi)^3} \int d^3q \ P_{{\cal R}_{\rm G}}(q)P_{{\cal R}_{\rm G}}(|\vec k-\vec q|)~, \label{oneloopchisq}
\end{align}
where $\mathcal{P}_{G^n}(k)\sim k^3\langle\mathcal{R}^n_{G}\mathcal{R}^n_{G}\rangle$. In the limit of infinite $\tilde{f}_{\rm NL}$, the parameter $\tilde{f}_{\rm NL}$ can be absorbed in the power spectrum amplitude but we write it explicitly to be clear with the notation. If we choose to characterize the curvature power spectrum as a Dirac-delta function, we can carry out an analytical integration of \cref{oneloopchisq}, which, after some steps, leads to the following expression 
\begin{align}\label{oneloopheavisidespectrum}
    \mathcal{P}_{G^2}(k) &= A^2_G\frac{k^2}{k^2_*}\theta_H (2 k_* - k)~,
\end{align}
with $\theta_H(k)$ being the heaviside step function. Notably, the spectrum is peaked at $k=2k_*$ and, it cuts off at $k>2k_*$ due to momentum conservation. The variance of the one-loop term is 
\begin{align}\label{oneloopvariance}
     \sigma_{\mathcal{R}_{G^2}}^2 &= \int\text{dln}\kt\frac{\mathcal{P}_{G^2}}{A_G^2}=2~.
\end{align}
The one-loop power spectrum \cref{oneloopheavisidespectrum} was used to construct the $\chi^2$-type non--Gaussian window function in \cite{Sharma:2024img} and study the spectral distortions in perturbative and non-perturbative limits. 

Hence, for the cubic case too, to determine the non--Gaussian corrections to spectral distortions from each diagram, we first calculate the variance of each distribution. With respect to $\mathcal{R}$, the variance for the dressed vertex term is given as
\begin{equation}
    \sigma_{\mathcal{R}_{G^3, DV}}^2=\int\text{dln}\kt\frac{\mathcal{P}_{G^3,DV}}{A_G^3}=9.
\end{equation}
For the sunset diagram term, we find
\begin{equation}
\sigma_{\mathcal{R}_{G^3, S}}^2=\int\text{dln}\kt\frac{\mathcal{P}_{G^3, S}}{A_G^3}=6.
\end{equation}
Thus, the variance of the full 2-loop power spectrum can be written as
\begin{align}
    \sigma_{\mathcal{R}^2_{G^3}} &= \sigma_{\mathcal{R}_{G^3, DV}}^2 +  \sigma_{\mathcal{R}_{G^3, S}}^2 \nonumber \\
    &= 9 A_G^3 + 6 A_G^3 \nonumber \\
    &= 15 A_G^3~.
    \label{eq:tot_var}
\end{align}
Next, we calculate the expected value of the comoving curvature perturbation, defined in \cref{R-expansion}. In our scenario, where we consider the dominant contribution to non--Gaussianity coming from the $\tilde{g}_\text{NL}$ term, we find
\begin{equation}
 \begin{split}
    \braket{\mathcal{R}\mathcal{R}}&= \braket{(\mathcal{R}_G+\tilde{g}_{\text{NL}}\mathcal{R}_G^3)(\mathcal{R}_G+\tilde{g}_{\text{NL}}\mathcal{R}_G^3)}\\   &=\braket{\mathcal{R}_G^2}+2\tilde{g}_{\text{NL}}\braket{\mathcal{R}_G^4}+\tilde{g}^2_{\text{NL}}\braket{\mathcal{R}_G^6}\\
    &=A_G+6\tilde{g}_{\text{NL}}A_G^2+15\tilde{g}^2_{\text{NL}}A_G^3.
 \end{split}\label{2loopExpected}
\end{equation}
However, if we go to the limit of pure cubic non--Gaussianity, which corresponds formally to taking the limit $g_\mathrm{NL} \rightarrow \infty$ and then absorbing $g_\mathrm{NL}$ in a redefinition of $A_G$, we only have the last term $\braket{\mathcal{R}\mathcal{R}}=15 A_G^3$, which matches (\ref{eq:tot_var}). Then, in consistency with \cite{Sharma:2024img, Byrnes:2024vjt}, we set $A=15 A_G^3$.

We arrive at the full power spectrum
\begin{equation}
    \mathcal{P}_{G^3}(k) = 3 A_G^3 \left[ \tilde{k}^3 p(\tilde{k}) + 3 k_* \delta(k-k_*) \right]~,
    \label{eq:full2loopbis}
\end{equation}
where we define the piecewise function
\begin{equation}\label{eq:p-definition}
p(\tilde{k})=\left\{ 
\begin{tabular}{ll}
$1$ & $\tilde{k}\leq 1$~, \\
$\frac{3-\tilde{k}}{2 \tilde{k}}$ & $1 < \tilde{k} < 3$~.
\end{tabular}
\right.
\end{equation}

\section{The $\mu$-distortion and PBH constraints}\label{ConstraintsSection}

\subsection{The spectral distortion constraints}

The $\mu$-distortion (or more generically $s$ distortion with $s=\mu, y$) in the limit of pure cubic non--Gaussianity follows from substituting ${\cal P}_{\cal R}(k)$ in Eq.~(3.22) of \cite{Sharma:2024img} with our new power spectrum from equation (\ref{eq:full2loopbis}). This gives 
\begin{equation}
    s = 9 A_G^3 \, W_s(k_*) + 6 A_G^3 \,W_s^{(CNG)}(k_*)
    = \frac{9}{15} A \, W_s(k_*) + \frac{6}{15} A \,W_s^{(CNG)}(k_*).
\end{equation}
Here, we use the same definition of the window function $W_s(k)$ as in \cite{Sharma:2024img}, 
\begin{equation}
W_s(k) = - \int dz \, {\cal I}_s(z)
    \frac{4 \dot{\tau}(z)\, S_\mathrm{ac}(k,z)}{(1+z)\,H(z)}~,
\end{equation}
and we define additionally
\begin{equation}
    W_s^\mathrm{(CNG)} (k_*)= 
    \frac{1}{2} \int_0^{3 k_*} \frac{dk}{k} \frac{k^3}{k_*^3} \,\, p(\tilde{k})
    \,\, W_s(k)~,
    \label{eq:wcng}
\end{equation}
where the superscript `CNG' stands for cubic non--Gaussianity.
For comparison, in \cite{Sharma:2024img}, we had
\begin{equation}
    W_s^\mathrm{(NG2)} (k_*)= 
    \frac{1}{2} \int_0^{2 k_*} \frac{dk}{k} \frac{k^2}{k_*^2} \,\,
    W_s(k)~,
    \label{eq:wng2}
\end{equation}
where the superscript `NG2' referred to $\chi^2$ (or quadratic) non--Gaussianity.
Note that equations \cref{eq:wcng,eq:wng2} amount to smoothing the window function $W_s(k)$ with two different kernels normalised to one.

\subsection{The PBH constraints}

We here follow the same technique for determining the PBH constraints on the power spectrum amplitude as introduced in \cite{Byrnes:2024vjt} and refer to that paper for details. For convenience we summarise a few key equations below.

We take the PBH mass to equal the horizon mass at the time the perturbation wavelength enters the horizon, and relate this to the comoving wavenumber $k$ via
\begin{equation}
    M_{\rm PBH}=M_H\simeq 17\left(\frac{g}{10.75}\right)^{-1/6}\left(\frac{k}{10^6\, {\rm Mpc}^{-1}}\right)^{-2} M_\odot, \label{M-k-relation}
\end{equation}
where $g$ is the number of relativistic degrees of freedom, which we approximate as equal to $10.75$ because we are interested in PBH formation at temperatures well below the QCD scale. We use Press Schechter theory to relate the power spectrum amplitude to the initial fraction of the universe which forms a PBH ($\beta$) and relate this in the standard manner to the fraction of PBHs in dark matter today, called $\fpbh\equiv\Omega_{\rm PBH}/\Omega_{\rm DM}$.

For Gaussian fluctuations, we use the following relation written in terms of the complementary error function
\begin{equation}
    \beta\simeq \frac12 \erfc\left(\frac{\calR_c}{\sqrt{A}}\right) \label{beta-GNL}
\end{equation} 
which is modified from the standard Press-Schechter result by a constant factor  because we include the intrinsic non-linear relation between $\delta$ and the curvature perturbation which remains even when $\mathcal{R}$ is Gaussian \cite{DeLuca:2019qsy,Young:2019yug,Gow:2020bzo}. We use this same multiplicative factor even when the curvature perturbations are intrinsically non--Gaussian, as done by e.g.~\cite{Unal:2020mts}. We choose a collapse threshold $\calR_c=0.67$ \cite{Nakama:2017qac} (note that working with the curvature perturbation directly should only be done for narrowly peaked power spectra). Our results are a good approximation to the much more sophisticated calculations in \cite{Gow:2020bzo}.
For the Gaussian cubed perturbations we follow the method developed in \cite{Byrnes:2012yx} to find that the variance is related to the collapse fraction $\beta_{{\rm G}^{3}}$ in terms of the inverse complementary error function by
\begin{equation}
    A_{G^3}=\frac{15}{8}\frac{\mathcal{R}_c^2}{\inverfc^3(2\beta_{{\rm G}^{3}})}\,.
\end{equation}

We stress that our simplified equations for relating $\beta$ to the power spectrum directly in terms of the curvature power spectrum can only be used for narrowly peaked power spectra. Our results also agree well with the more complex approach followed by \cite{Wang:2025hbw}.

Given that SMBHs can only be a small fraction of the total dark matter density, we show constraints for $\fpbh=10^{-5}$ and $10^{-10}$. We neglect the impact of accretion given the uncertainty in modelling it. In the extreme case that all PBHs had increased their mass by a factor of $10^5$ the $\fpbh=10^{-5}$ constraint (based on the PBH mass today) should really be compared to the constraint on the primordial power spectrum variance of $\fpbh=10^{-10}$ (based on their mass at formation) combined with an appropriate shift in the mass to $k$ relation.

We also include the constraint of having zero PBHs inside the entire observable cosmological horizon, which is 
\begin{equation}    \beta<\betaz(k)\simeq1.2\times10^{-11}(k \, {\rm Mpc})^{-3}, \label{1-pbh}
\end{equation}    
where $\betaz$ denotes the PBH fraction at formation with one PBH in today's cosmological horizon \cite{Carr:1997cn,Cole:2017gle,Nakama:2017xvq}. When this PBH constraint on the power spectrum is tighter than the spectral distortion constraint no PBHs can be generated. We will see that this is the case for a large range of PBH masses if the perturbations are Gaussian or $\chi^2$ distributed, which is one motivation for studying Gaussian cubed perturbations.

\subsection{The constraint plots}\label{constraintplotsection}

\begin{figure}[!ht]
    \centering
    \includegraphics[width=\linewidth]{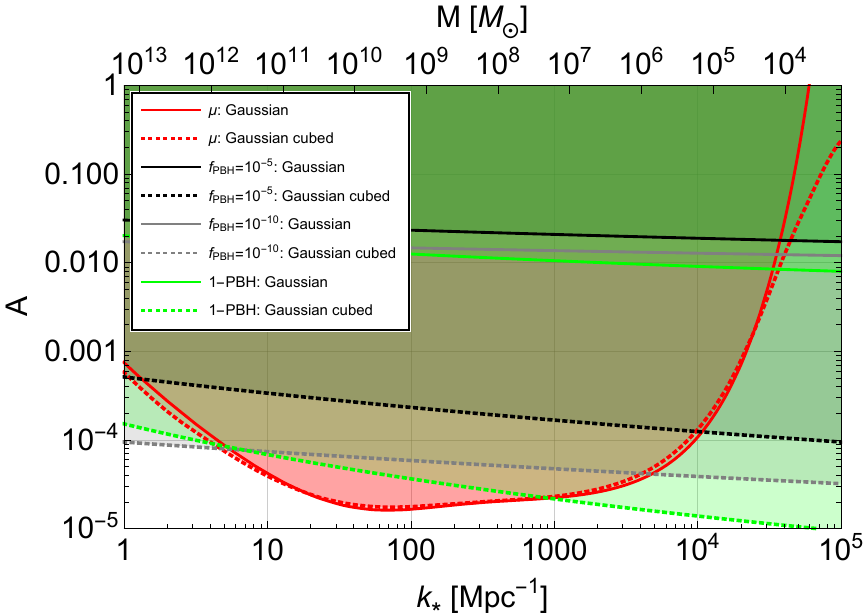}
    \caption{The $\mu$-distortion constraints on the power spectrum variance for a Gaussian perturbation distribution (solid lines), as well as a Gaussian cubed distribution (dotted lines). We also show the power spectrum amplitude required to generate several different abundances of PBHs, for both of the perturbation distributions.}
    \label{fig:G3-G}
\end{figure}

In \cref{fig:G3-G}, we show how the assumed density perturbation distribution impacts both the $\mu$-distortion constraints and the power spectrum amplitude required to generate a given abundance of PBHs. Specifically, we compare a purely Gaussian and a Gaussian cubed distribution, as detailed throughout the paper. As can be seen in the figure, there is little variation between the $\mu$-distortion constraints. The departure from Gaussian constraints becomes apparent only in the tail at larger $k$ (i.e. for $k\gtrsim5\times10^{4}\text{Mpc}^{-1}$). This allows us to strengthen the conclusions of \cite{Sharma:2024img, Byrnes:2024vjt}, and show that $\mu$ constraints are robust even in the extreme case of infinite $g_\text{NL}$.\\

Furthermore, the fact that the $\mu$ constraints are similar in the respective cases deters us from investigating the case of perturbative $g_\text{NL}$ non--Gaussianity. Firstly, the $\mu$ constraint curve expected from a perturbative non--Gaussian approach will lie somewhere in between the two curves plotted in \cref{fig:G3-G}. Thus, the constraint will be even closer to that of the Gaussian. Secondly, the abundance curves for each PBH scenario will shift upwards, meaning a larger power spectrum amplitude will be required to generate the same abundance. Overall, the perturbative approach is less exciting and one could predict the conclusions without the need of it's own study.\\

\begin{figure}[!ht]
    \centering
    \includegraphics[width=\linewidth]{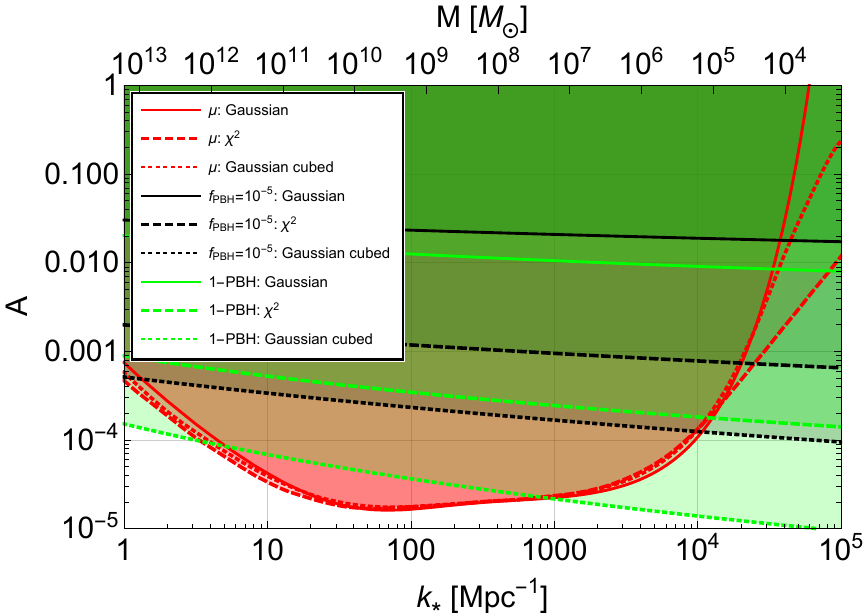}
    \caption{Constraints on the power spectrum variance for a Gaussian perturbation (solid lines), Gaussian squared perturbation (dashed) or Gaussian cubed perturbation (dotted lines). Note that the $f_{\rm PBH}=10^{-10}$ constraint has been removed to reduce clutter.}
    \label{fig:G3-G2-G}
\end{figure}

In \cref{fig:G3-G2-G}, we additionally consider the case of $\chi^2$ non--Gaussianity. Regarding the $\mu$-distortion constraints, we see an interesting feature of increasing the order in the $\mathcal{R}_G$ Taylor expansion. Namely that, in the high-$k$ tail, the Gaussian cubed distribution constraints resemble the Gaussian constraints more than that of the $\chi^2$. As discussed in \cref{pureSunsetSection}, this is expected. Regarding the PBH abundances, in the Gaussian cubed case the unconstrained power spectrum amplitude required to generate a single PBH in today’s cosmological horizon is an order of magnitude less than that of $\chi^2$. In other words, forming a single PBH of mass $\mathcal{O}(10^7M_\odot)$ is possible in the Gaussian cubed case, whereas in the $\chi^2$ case the maximum mass is $\mathcal{O}(10^5M_\odot)$. This difference becomes less apparent as you increase the PBH abundance. 

Importantly, we have shown that, despite the order of non--Gaussianity, spectral distortion constraints on the power spectrum amplitude remain a large obstacle in describing any population of primordial SMBHs. In \cite{Sharma:2024img, Byrnes:2024vjt}, we made the prediction that, regardless of the order in non--Gaussianity, there would still be a range in $k$ in which the window function dominates. In this range, we expect the power spectrum shape to have a minimal impact on the spectral distortion constraints. These expectations have been strengthened by \cref{fig:G3-G2-G}, where we clearly see strong alignment of $\mu$-constraint curves in the range ($10\text{Mpc}^{-1}\lesssim k\lesssim10^4\text{Mpc}^{-1}$). This is discussed in more detail in \cref{pureSunsetSection}.



\section{Conclusions}\label{conclusion}

Supermassive black holes have been suggested to lie in the centers of galaxies since the discovery of quasars
\cite{Schmidt:1963wkp, 1963Natur.197.1041G}. The formation of such objects is still unclear, and an active area of research \cite{Cammelli:2024uhh,Liempi:2024qgg,2025A&A...695A..97D,Mone:2024yxl,Chen:2025onr}. Primordial black holes, typically thought to form with the horizon mass of the universe at the time of formation, are able to avoid many of the difficulties associated with astrophysical formation. However, on the mass scale associated with supermassive black holes, the primordial curvature power spectrum is tightly 
constrained by spectral distortions. Consequently, it is challenging to motivate supermassive black holes of primordial origin. 

It has been suggested that, if the density perturbations follow a highly non--Gaussian distribution, these spectral distortion constraints could be evaded \cite{Nakama:2016kfq,Nakama:2017xvq,Hooper:2023nnl}. This has been the topic of several recent studies, with a focus on local-type non--Gaussianity \cite{Sharma:2024img, Byrnes:2024vjt,Hai-LongHuang:2024vvz}. In this work we further develop the argument that, even with large non--Gaussianity, spectral distortion constraints are unavoidable. Specifically, we have shown that even in the case of infinite $\gnl$ supermassive black holes cannot be primordial. Our arguments follow the calculation of a 2-loop integral, which we perform analytically. The result is given by \cref{eq:full2loopbis}. Furthermore, we have made the first constraints on the curvature power spectrum using spectral distortions in this context. These constraints are plotted and discussed in \cref{constraintplotsection}.

We are also able to make a stronger argument that the distortion constraints on higher-order non--Gaussian perturbations will remain similar to the Gaussian constraints, especially on the scales where the constraints are tightest which corresponds to the most interesting supermassive black hole range. We detail the argument in \cref{infiniteLoopSection} and in particular show that the only diagrams for which we cannot calculate the explicit scale dependence for $n=4$ and $n=5$ correspond to a 25\% and 13\% correction respectively to the total variance. Following the results for the power spectrum amplitude required for PBH generation derived in \cite{Byrnes:2024vjt} (see Fig.~6 in that paper) we therefore remain confident that at least a small number of primordial SMBHs can be generated with $n=4$ and a value as large as $\fpbh=10^{-5}$ if $n=5$. However, we caution that there remains uncertainty in the determination of the PBH abundance, especially with non-perturbative non--Gaussianity, and that other observational challenges remain as summarised in \cite{Byrnes:2024vjt} (for recent additions on the question of PBH clustering see \cite{Kasai:2024tgu,Huang:2024aog}). \\ \\
{\bf Acknowledgements} CB thanks David Seery for insightful discussions about the meaning of the different loop diagrams and Hai-Long Huang for discussions about the results in \cite{Hai-LongHuang:2024vvz}. XP thanks Tom Gent for helpful discussion. XP is supported by an STFC studentship. CB is supported by STFC grants ST/X001040/1 and ST/X000796/1. DS is supported by a grant from the German Academic Exchange Service (DAAD). 

\appendix\label{appendix}

\section{Detailed solution of the 2-loop sunset diagram}\label{detailedSection}
In this appendix we add details to the solution of the convolution integrals sketched in \cref{sunsetsection} leading to our key analytic result \cref{eq:full2loopbis}.

\subsection{$\theta_3$ derivation}\label{sec:theta3}
In order to calculate each 2-loop diagram, we are essentially interested in 3 vectors. In the case of the dressed vertex diagram, these vectors are independent of each other. However, in the case of the sunset diagram, the angles between each of these vectors are nontrivial. As seen in the main text, the result is a much more involved calculation. Two of the integration variables are $\theta_1$ and $\theta_2$, which correspond to the angle made with $\vec{k}$ between $\vec{q_1}$ and $\vec{q_2}$, respectively. The angle between $\vec{q_1}$ and $\vec{q_2}$, denoted as $\theta_3$, then depends on $\theta_1$ and $\theta_2$.

We derive the $\theta_3$ relation using the transformation from Cartesian to spherical polar coordinates
$$ x=r\sin\theta\cos\phi, \qquad y=r\sin\theta\sin\phi, \qquad z=r\cos\theta. $$
The 2 unit vectors, $\hat{q_1}$ and $\hat{q_2}$, written in this coordinate system are
$$ \hat q_1=(1,\theta_1,0), \qquad \hat q_2=(1,\theta_2,\phi) $$
Then, the dot product is given by
$$\cos\theta_3\equiv\hat q_1\cdot \hat q_2 =\sin\theta_1\sin\theta_2\cos\phi+\cos\theta_1\cos\theta_2. $$

\subsection{Arccos solutions}\label{sec:arcos}

\begin{figure}[ht!]
    \centering
    \includegraphics[width=\linewidth]{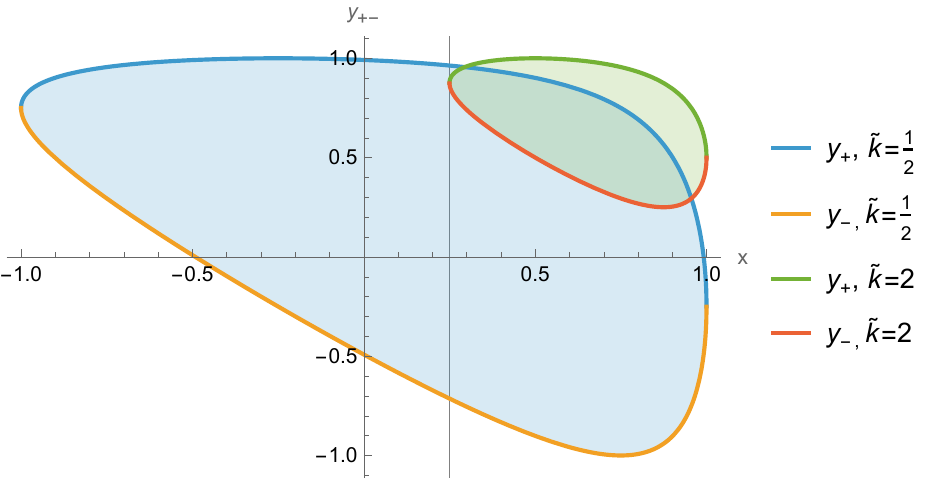}
    \caption{We show the region of integration defined by $y_\pm$ and $x_\pm$, for 2 different values of $\kt$. As can be seen, for $\kt<1$ the $x_\pm$ limits are unchanged from $\pm1$, whereas for $\kt>1$ the lower $x$ integration limit is altered. Accordingly, we plot the vertical line corresponding to $\frac{\kt^2-3}{2\kt}$ evaluated at $\kt=2$.}
    \label{IntegrationArea}
\end{figure}

In the integration range of interest $F(\phi)$, as defined in \cref{DiracDelta}, has solutions
\begin{equation}
 \phi_j=\text{Arccos}\left(\frac{g(x, y, \kt)}{h(x, y, \kt)}\right), \hspace{4mm}2\pi-\text{Arccos}\left(\frac{g(x, y, \kt)}{h(x, y, \kt)}\right),
\end{equation}
where
\begin{equation}
    \begin{split}
        &g(x, y, \kt)=2\kt(x+y)-\kt^2-1-2xy,\\
        &h(x, y, \kt)=2\sqrt{(1-x^2)(1-y^2)}.
    \end{split}
\end{equation}
Requiring that these solutions are real is equivalent to imposing
\begin{equation}\label{inequality}
    h^2(x, y, \kt)-g^2(x, y, \kt)\geq0.
\end{equation}
Written explicitly, this is
\begin{equation}
    4(1-x^2)(1-y^2)-(2\kt(x+y)-\kt^2-1-2xy)^2\geq0.
\end{equation}
This inequality, whose solutions will define the subsequent limits of integration, can be written in the form
\begin{equation}
    \alpha y^2+\beta y +\gamma\geq0,\label{Quadranequality}
\end{equation}
with coefficients
\begin{equation}
    \begin{split}
        &\alpha= -4 (1 + \kt^2 - 2 \kt x)\\
        &\beta=  4 (\kt - x) (1 + \kt^2 - 2 \kt x)\\
        &\gamma= 3 - \kt^4 + 4 \kt x + 4 \kt^3 x - 4 x^2 - 2 \kt^2 (1 + 2 x^2).
    \end{split}
\end{equation}
We next define the derivative of $F(\phi)$ in terms of $g$ and $h$
\begin{equation}
 \begin{split}
    F'(\phi)&=\frac{k_*\sqrt{
    (1-x^2)(1-y^2)}\text{sin}(\phi)}{\sqrt{\kt^2-2\kt(x+y)+2(1+xy+\sqrt{(1-x^2)(1-y^2)}\text{cos}(\phi))}}\\
    &=\frac{k_*h(x, y, \kt)\text{sin}(\phi)}{2\sqrt{1-g(x, y, \kt)+h(x, y, \kt)\text{cos}(\phi)}}.
 \end{split}
\end{equation}
Then, it is straightforward to see
\begin{equation}\label{fgFprime}
 \begin{split}
    F'(\phi_j)&=\frac{k_*h(x, y, \kt)\sqrt{1-\frac{g^2(x, y, \kt)}{h^2(x, y, \kt)}}}{2\sqrt{1-g(x, y, \kt)+h(x, y, \kt)\frac{g(x, y, \kt)}{h(x, y, \kt)}}}\\
    &=\frac{k_*}{2}\sqrt{h^2(x, y, \kt)-g^2(x, y, \kt)}
    =\frac{k_*}{2}\sqrt{\alpha y^2 +\beta y +\gamma},
 \end{split}    
\end{equation}
where we have used the following equality 
\begin{equation}
    \text{sin}(\text{arccos}(\phi))=\text{sin}(\lambda)=\sqrt{1-\text{cos}^2(\lambda)}=\sqrt{1-(\text{cos}(\text{arccos}(\phi))^2}=\sqrt{1-\phi^2}.
\end{equation}
As can be seen, the coefficients of $y$ are functions of $x$, leading to some modifications of the integration limits of both variables following the integral of $\phi_1$. The integration limits for $y$ are given as
\begin{equation}\label{yplusminus}
    y_\pm(\kt,x)=\frac{(\kt-x)}{2}\pm\frac{1}{2}\sqrt{\frac{(x^2-1)(-3+\kt^2-2\kt x)}{1+\kt^2-2\kt x}},
\end{equation}
which we show, for some example choices of $\kt$, in \cref{IntegrationArea}.\\

As discussed in the text, our integration of the sunset diagram should be thought of within two ranges of $\kt$. This can be seen by looking at each term within our $y_\pm$ solution. Specifically, within the square root. Firstly, the $(x^2-1)$ term is always negative, whereas $1+\kt^2-2\kt x$ is always positive. Then, for $\kt<1$, we find that $-3+\kt^2-2\kt x$ is always negative. Thus, the square root term is always real. However, for $\kt>1$, this term obtains a zero at $x=\frac{\kt^2-3}{2\kt}$ such that, for $x<\frac{\kt^2-3}{2\kt}$, the term is positive. The result is an imaginary integration limit, which is unphysical. Thus, we change the lower $x$ integration limit to $\frac{\kt^2-3}{2\kt}$. 

Following the above argument, we examine the condition for an analytic ansatz to \cref{arcsinsolution}. That is, that the discriminant, $\Delta$, is negative. Written explicitly, we see
\begin{equation}\label{discriminant}
    \Delta=-16  (-1 + x^2) (-3+\kt^2-2\kt x)(1+\kt^2-2\kt x).
\end{equation}
Thus, we find that requiring analyticity is the exact same condition as requiring the square root argument in our $y_\pm$ integration region to be positive. The result is that, for this diagram, we have a complete analytic solution for all $\kt$ values. 

\subsection{A simplified arcsin solution}\label{sec:arcsin}

After plugging in our integration limits into the integral with respect to $y$, \cref{arcsinsolution}, we see that
\begin{equation}
 \begin{split}
I&=\frac{-1}{\sqrt{-\alpha}}\text{arcsin}\left( \frac{2\alpha y_\pm+\beta}{\sqrt{-\Delta}} \right),\\
&=\frac{-1}{\sqrt{-\alpha}}\text{arcsin}\left( \frac{2\alpha\left(\frac{-\beta\mp\sqrt{\beta^2-4\alpha\gamma}}{2\alpha}\right)+\beta}{\sqrt{\beta^2-4\alpha\gamma}}\right),\\
&=\frac{-1}{\sqrt{-\alpha}}\text{arcsin}\left(\mp1\right)=-(\frac{-\pi}{2\sqrt{-\alpha}}-\frac{\pi}{2\sqrt{-\alpha}})=\frac{\pi}{\sqrt{-\alpha}}.
 \end{split}
\end{equation}

The reason we use $\mp$ when we plug the $y_\pm$ solutions into our integration limits is subtle. This can be seen by plugging in our coefficients into the quadratic formula solution for $y$. We see the following
\begin{equation}
 \begin{split}
    y&=\frac{-\beta\pm\sqrt{\beta^2-4\alpha\gamma}}{2\alpha}\\
    &=\frac{-4(\kt-x)(1+\kt^2-2\kt x)\pm4\sqrt{(x^2-1)(1+\kt^2-2\kt x)(-3+\kt^2-2\kt x})}{-8(1+\kt^2-2\kt x)}\\
    &=\frac{(\kt-x)}{2}\mp\frac{1}{2}\sqrt{\frac{(x^2-1)(-3+\kt^2-2\kt x)}{1+\kt^2-2\kt x}}.
 \end{split}    
\end{equation}
From this and \cref{yplusminus}, we can clearly see that our $y_\pm$ solutions can be written as
\begin{equation}
    y_\pm=\frac{-\beta\mp\sqrt{\beta^2-4\alpha\gamma}}{2\alpha}.
\end{equation}

\section{Spectral distortion constraints from the sunset diagram alone}\label{pureSunsetSection}

\begin{figure}[!ht]
    \centering
    \includegraphics[width=0.8\linewidth]{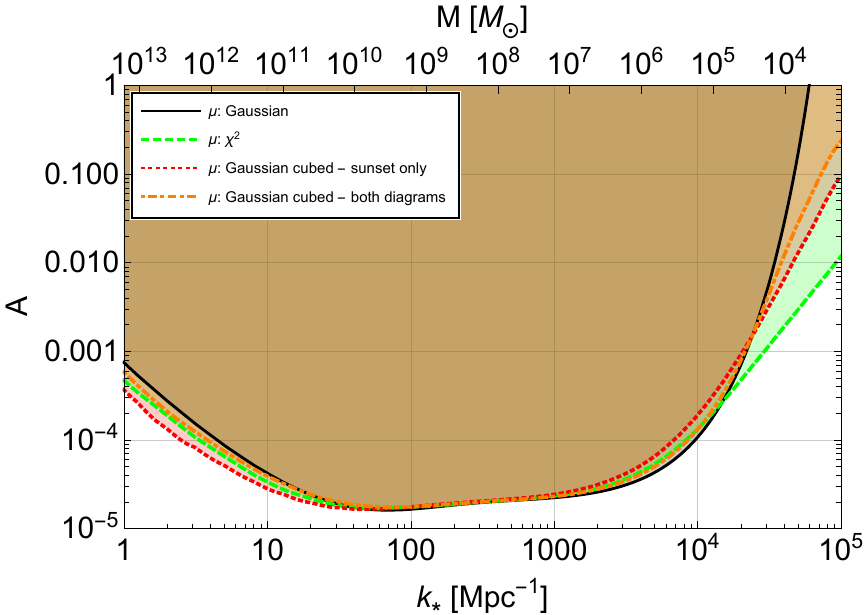}
    \caption{Spectral $\mu$-distortion constraints on the power spectrum variance for a Gaussian perturbation (solid), $\chi^2$ (dashed) or the Gaussian cubed perturbation with only the sunset diagram contribution (short dashes) and the full contribution from both diagrams (dot-dashed lines).}
    \label{fig:mu}
\end{figure}

As we have discussed and shown in \cref{DiagramPlot1} there are two diagrams which contribute to the power spectrum when considering a Gaussian cubed perturbation. Up to here we have always treated both diagrams equally, but for completeness we here also show the constraint which arises if one only considers the sunset diagram in \cref{fig:mu}.

\begin{figure}[!ht]
    \centering
    \includegraphics[width=0.9\linewidth]{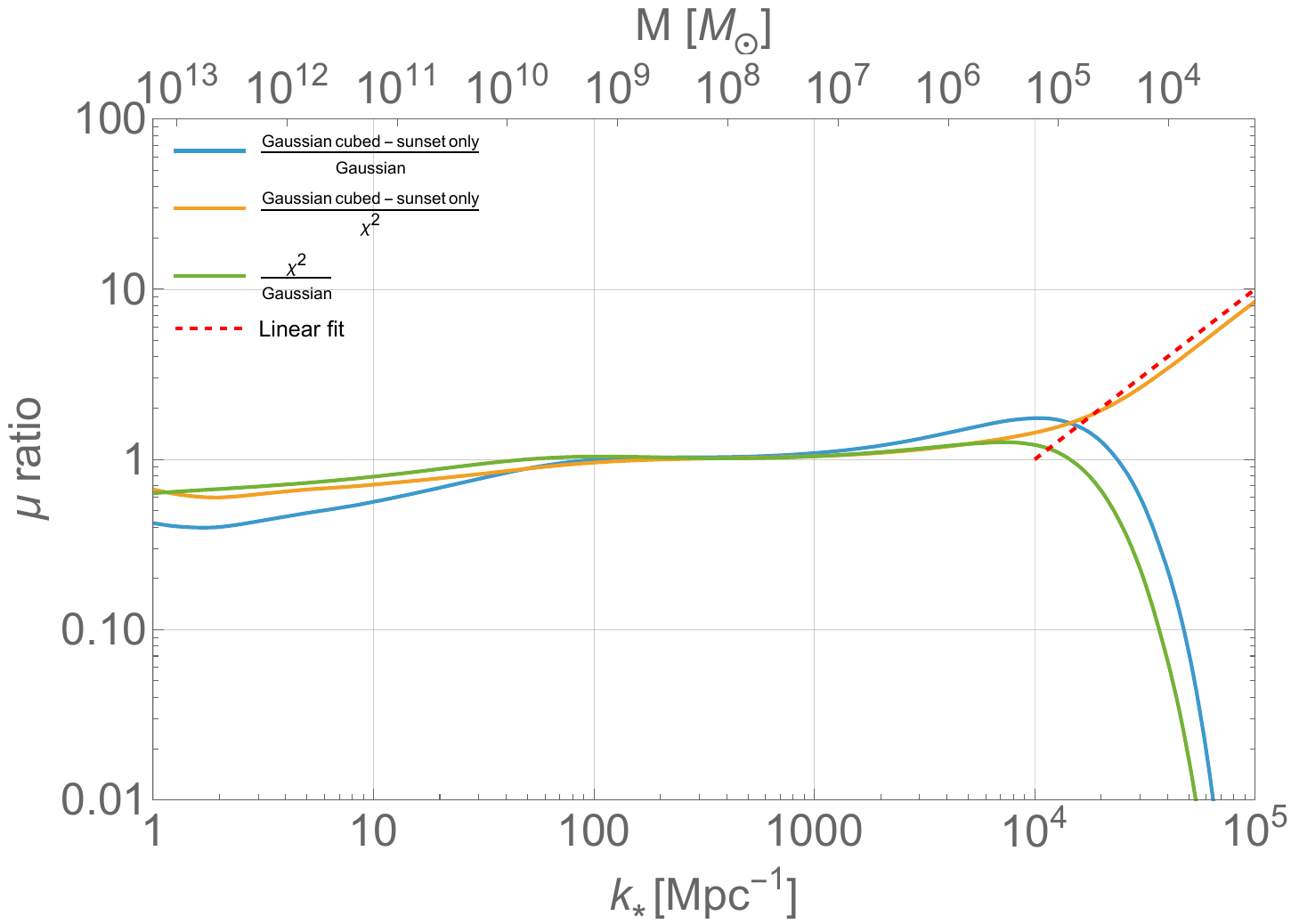}
    \caption{This figure compares the ratio of the $\mu$ amplitudes for Gaussian cubed and $\chi^2$ non-Gaussianity, as well as the Gaussian case. The linear fit shows the expected relation between the constraints at high-$k$.}
    \label{fig:muRatio}
\end{figure}

This is equivalent to taking the power spectrum as \cref{eq:full2loopbis} but without the Dirac-delta function contribution. As expected, we can see that the result deviates slightly more strongly from the Gaussian constraint line than the full result, because the contribution to the power spectrum from the dressed vertex diagram has the same (Dirac-delta) shape as the Gaussian power spectrum. Towards the tail for larger values of $k_*$, both Gaussian cubed constraints are significantly more similar to the Gaussian constraint than the $\chi^2$ constraint, which is due to the fact that the $\chi^2$ power spectrum has a flatter slope ($k^2$ instead of $k^3$) towards small values of $k$. In \cref{fig:muRatio} we show the ratio of the constraints which show that towards large $k$ the ratio is indeed proportional to $k$. 

One motivation for calculating the constraint with the power spectrum shape coming exclusively from the sunset diagram is that \cite{Byrnes:2007tm} speculated that it may be possible to renormalise away the dressed vertex diagrams. Whilst this (in a more general context) is now a hot topic which is still not resolved, see e.g.~\cite{Iacconi:2023ggt, Kristiano:2024ngc} for an introduction and references therein, in the extreme case of infinite $\gnl$ which we are considering there are no other diagrams for the power spectrum at any loop order. We therefore do not expect that the dressed vertex diagram should be neglected, but include this appendix for completeness. The setting sun diagram may be interpreted as non-linear shot noise arising from small scale modes, which therefore decreases on large scales, $k\ll k_*$, and has a cut-off on small scales. The dressed vertices can be seen as the effective generation of a power spectrum with the same shape as the Gaussian power spectrum, albeit only existing at loop-level in our case and hence higher order. We remind the reader that we are working in the simplified case that the `input' perturbations are purely Gaussian cubed with a delta--function power spectrum and that we are only calculating `classical' loop diagrams relating to the $\delta$-$N$ formalism as explained in \cite{Byrnes:2007tm}. Our interpretation of the diagrams may not apply in more general situations.

\section{The extension to infinite--loop orders}\label{infiniteLoopSection}

We have developed a method to compute each diagram's contribution to the non--Gaussian power spectrum variance; see \cref{VarianceSection}. This can straightforwardly be applied to higher-order non--Gaussianity assuming the nth-order term is the dominant one, which we note requires fine-tuning \cite{Boubekeur:2005fj,Nelson:2012sb,Nurmi:2013xv,Young:2014oea}. In doing so, we find the sunset diagram becomes more subdominant as you increase the order of pure non--Gaussianity. Firstly, in a similar fashion to \cref{R-expansion}, we calculate the $n$-loop variance of the curvature perturbation. For even $n$ this is given by
\begin{equation}
 \begin{split}
    \braket{\mathcal{R}\mathcal{R}}&= \braket{(\mathcal{R}_G+\tilde{n}_{\text{NL}}(\mathcal{R}_G^n-\braket{\mathcal{R}_G^n}))(\mathcal{R}_G+\tilde{n}_{\text{NL}}(\mathcal{R}_G^n-\braket{\mathcal{R}_G^n}))}\\  &=\braket{\mathcal{R}_G^2}+\tilde{n}^2_{\text{NL}}\braket{(\mathcal{R}_G^n-\braket{\mathcal{R}_G^n})^2},\\
 \end{split}
\end{equation}
where $\braket{\mathcal{R}_G^2}=A_G$, $\tilde{n}_{\text{NL}}$ is the normalised coefficient of the $n$-th term in the Taylor expansion and we have used the fact that the expectation values of odd powers of the Gaussian curvature perturbation are 0. The corresponding calculation for odd $n$ is
\begin{equation}
 \begin{split}
    \braket{\mathcal{R}\mathcal{R}}&= \braket{(\mathcal{R}_G+\tilde{n}_{\text{NL}}\mathcal{R}_G^n)(\mathcal{R}_G+\tilde{n}_{\text{NL}}\mathcal{R}_G^n)}\\  &=\braket{\mathcal{R}_G^2}+2\tilde{n}_{\text{NL}}\braket{\mathcal{R}_G^{n+1}}+\tilde{n}^2_{\text{NL}}\braket{\mathcal{R}_G^{2n}}.\\ 
 \end{split}
\end{equation}
Taking the limit of infinite $\tilde{n}_{\text{NL}}$, the term proportional to $\tilde{n}_{\text{NL}}^2$ will dominate. Similarly to \cref{VarianceSection}, we can absorb the term proportional to $\tilde{n}_{\text{NL}}$ in a redefinition of the amplitude $A_G$. Then, we find that the variance for all $n$ can be written
\begin{equation}
\braket{\mathcal{R}\mathcal{R}}=
      \braket{\mathcal{R}_G^{2n}}-\braket{\braket{\mathcal{R}_G^n}^2},
\end{equation}
where the second term is 0 for odd $n$.

\begin{figure}[!ht]
    \centering
    \includegraphics[width=0.7\linewidth]{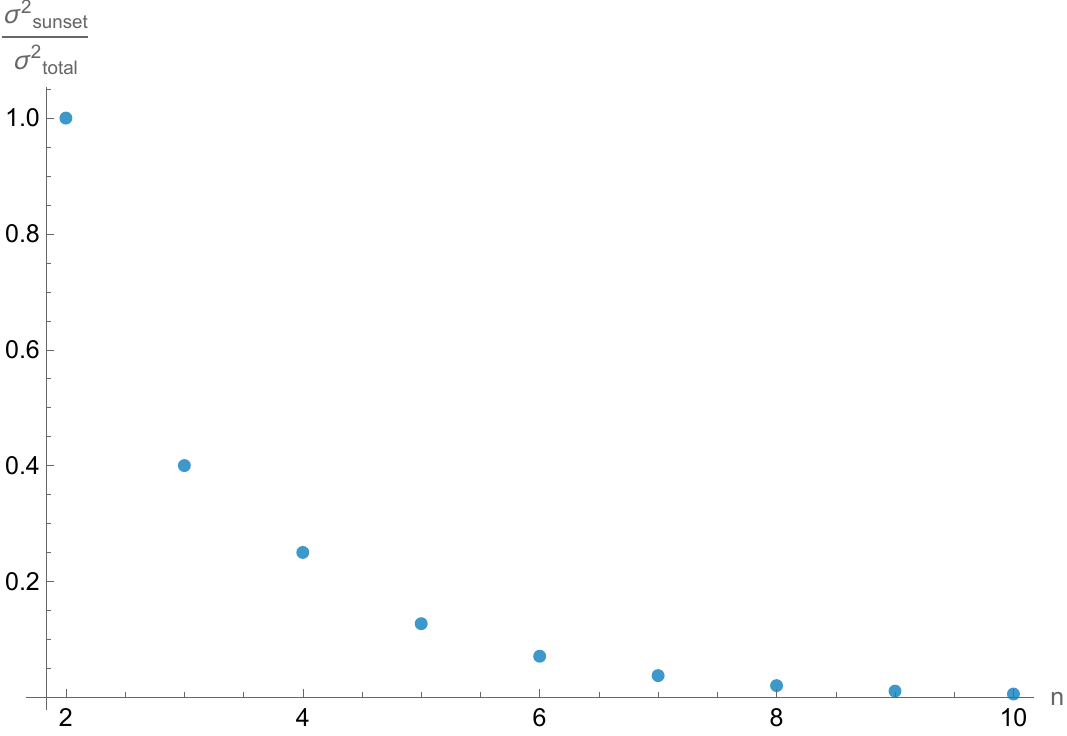}
    \caption{The contribution of the (pure) sunset diagram to the overall power spectrum variance, with non--Gaussian corrections as a function of the order of the non--Gaussianity.}
    \label{fig:ComplicatedContribution}
\end{figure}

Following \cite{Byrnes:2007tm}, the prefactor of the higher order equivalent to the 2-loop sunset diagram is given by the factorial of $n$. Then, we see that the overall contribution can be expressed as 
\begin{equation}
    \frac{n!}{\braket{\mathcal{R}\mathcal{R}}}=\frac{n!}{\braket{\mathcal{R}_G^{2n}}-\braket{\braket{\mathcal{R}_G^n}^2}}=\frac{n\Gamma[n]^2\Gamma[n/2]^2}{2^{1-n}(\Gamma[2n]\Gamma[n/2]^2-(1+(-1)^n)\Gamma[n]^3)},
\end{equation}
where $n$ is the order of the non--Gaussianity, and we have used the relation \cite{2012arXiv1209.4340W}
\begin{equation}
    \braket{\mathcal{R}_G^n}=\frac{1}{\sqrt{\pi}}\left(2^{-1+\frac{n}{2}}(1-(-1)^n)\Gamma\left(\frac{1+n}{2}\right)\right),
\end{equation}
where $\Gamma[n]=(n-1)!$. We plot this contribution in \cref{fig:ComplicatedContribution}.

Furthermore, we provide additional details of the corresponding n-loop power spectrum. Firstly, all diagrams are symmetric between the two vertices and can be of the form of dressed vertex diagrams, sunset diagrams or a combination of both (with the combination diagram first arising for $n=4$). The rules are given in \cite{Byrnes:2007tm} using a different notation and are summarised here for convenience. The pure sunset diagram (meaning one without any dressed vertices) in general is given by
\begin{equation} \label{eq:sunset-general}
    P_{\rm Sunset}^{(n-1)-{\rm loop}}=n!\int \left(\prod\limits_{i=1}^{n-1}\frac{d^3q_i}{(2\pi)^3} P_G(q_i)\right)P_G(|\vec{k}-\sum\limits_{i=1}^{n-1}q_i|).
\end{equation}
The corresponding $f$ of \cref{f-kstar} to the above integral is given as
\begin{equation}
   f_{(n-1)-\text{loop}} = k_*\sqrt{\kt^2-2\kt\sum\limits_i^{n-1}x_i+n-1+\sum\limits_{j\neq i}\sum\limits_{i=1}^{n-1}(x_ix_j+\sqrt{(1-x_i^2)(1-x_j^2)}\text{cos}(\phi_i-\phi_j))},
\end{equation}
where we can choose $\phi_1=0$. Thus, for $n>3$, there is an additional type of term of the form $\text{cos}(\phi_i-\phi_j)$ with $\phi_i,\phi_j\neq0$. This makes the solution of \cref{DiracDelta} much more difficult to solve. Furthermore, it is unlikely that the integral of the corresponding $|F_n'(\phi_j)|^{-1}$ will have a known analytic solution.

\subsection{Semi-analytic expression for $n=4$ (Gaussian to power 4)}

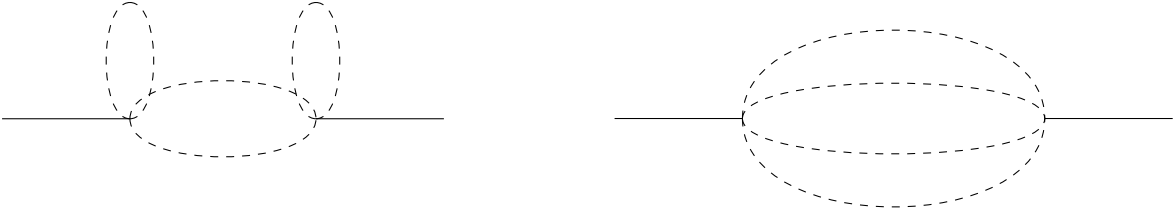
\begin{figure}[ht!]
  \begin{minipage}[b]{0.5\textwidth}
   \begin{tikzpicture}[scale=3.075, transform shape]
      \begin{feynman}
        \vertex (a); 
        \vertex [right = 0.55cm  of a] (b);
        \vertex [below = 0.506cm of a] (a1);
        \vertex [right = 0.8cm  of b] (c);
        \vertex [right = 0.55cm  of c] (d);
        \vertex [above = 0.5cm  of b] (e);
        \vertex [above = 0.5cm  of c] (f);

        \diagram* {(a) -- [] (b), 
          (a) -- [opacity=0] (a1),
          (b) -- [scalar, half right, looseness=0.7] (e), 
          (b) -- [scalar, half left, looseness=0.7] (e), 
           (b) -- [scalar, half left, looseness=0.7](c),
          (b) -- [scalar, half right, looseness=0.7] (c), 
          (c) -- [scalar, half right, looseness=0.7] (f), 
          (c) -- [scalar, half left, looseness=0.7] (f), 
          (c) -- [] (d),
          };
      \end{feynman}
\end{tikzpicture}
  \end{minipage}
  \hspace{2mm} 
  \begin{minipage}[b]{0.5\textwidth}
   \begin{tikzpicture}[scale=3.075, transform shape]
       \begin{feynman}
        \vertex (a); 
        \vertex [right = 0.55cm  of a] (b);
        \vertex [right = 1.3cm  of b] (c);
        \vertex [right = 0.55cm  of c] (d);

        \diagram* {
          (a) -- [] (b), 
          (b) -- [scalar, half right, looseness=0.4] (c), 
          (b) -- [scalar, half left, looseness=0.4](c),
          (b) -- [scalar, half right, looseness=1] (c), 
          (b) -- [scalar, half left, looseness=1](c),
          (c) -- [] (d)
          };
      \end{feynman}
\end{tikzpicture}
  \end{minipage}
  \caption{The loop terms for $n=4$. {\it Left}: The simpler term, which is straightforward to calculate given the results of the 1-loop calculation. {\it Right}: The more complicated 3-loop ``sunset'' integral.}\label{DiagramPlot2}
\end{figure}

For this case, there are again two diagrams, and each are 3-loop diagrams. We show these diagrams in \cref{DiagramPlot2}. One is a mixed diagram of the form of the 1-loop sunset diagram combined with each vertex being dressed whose explicit form is given by
\begin{equation}
    P_{\rm mixed}=\frac{(4!)^2}{2^2\times2} \prod\limits_{i=1}^{2}\left( \frac{d^3q_i}{(2\pi)^3} P_G(q_i)\right)\int \frac{d^3q_3}{(2\pi)^3} P_G(q_3)P_G(|\vec{k}-\vec{q_3}|)
\end{equation}
and using the analytic solution for $\chi^2$ non--Gaussianity \cref{oneloopheavisidespectrum} we can determine the following analytic expression
\begin{equation}
   {\mathcal P}_{\rm mixed}=  36 A^4_G\frac{k^2}{k^2_*}\theta_H (2 k_* - k)
\end{equation}
whose variance is 72 $A^4_G$.

The pure sunset diagram has a $k$ dependence formally given by the convolution integral of \cref{eq:sunset-general} with variance $96-72 A^4_G=24 A^4_G=4! A^4_G$. Therefore, the relative contribution of this diagram to the total variance is $1/4$ as shown in \cref{fig:ComplicatedContribution}.

\subsection{Semi-analytic expression for $n=5$ (Gaussian to power 5)}

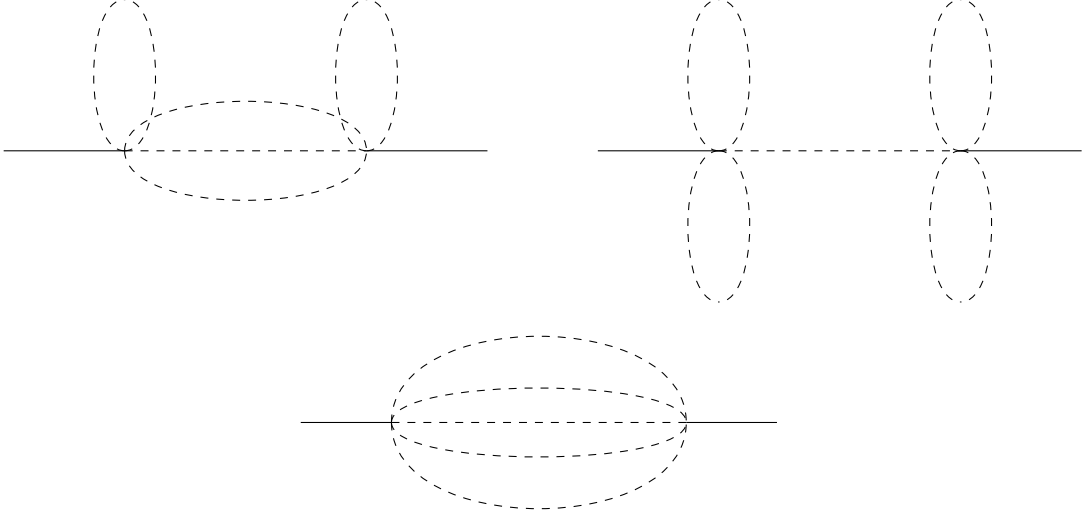
\begin{figure}[ht!]
\begin{minipage}[b]{0.5\linewidth}
\begin{tikzpicture}[scale=4, transform shape]
\begin{feynman}
        \vertex (a); 
        \vertex [right = 0.4cm  of a] (b);
        \vertex [below = 0.506cm of a] (a1);
        \vertex [right = 0.8cm  of b] (c);
        \vertex [right = 0.4cm  of c] (d);
        \vertex [above = 0.5cm  of b] (e);
        \vertex [above = 0.5cm  of c] (f);

        \diagram* {(a) -- [] (b), 
          (a) -- [opacity=0] (a1),
          (b) -- [scalar, half right, looseness=0.7] (e), 
          (b) -- [scalar, half left, looseness=0.7] (e), 
           (b) -- [scalar, half left, looseness=0.7](c),
           (b) -- [scalar, half left, looseness=0](c),
          (b) -- [scalar, half right, looseness=0.7] (c), 
          (c) -- [scalar, half right, looseness=0.7] (f), 
          (c) -- [scalar, half left, looseness=0.7] (f), 
          (c) -- [] (d),
          };
\end{feynman}
\end{tikzpicture}
 \end{minipage}
  \begin{minipage}[b]{0.4\textwidth}
   \begin{tikzpicture}[scale=4, transform shape]
   \begin{feynman}
        \vertex (a); 
        \vertex [right = 0.4cm  of a] (b);
        \vertex [below = 0.506cm of a] (a1);
        \vertex [right = 0.8cm  of b] (c);
        \vertex [right = 0.4cm  of c] (d);
        \vertex [above = 0.5cm  of b] (e);
        \vertex [above = 0.5cm  of c] (f);
        \vertex [below = 0.5cm  of b] (e1);
        \vertex [below = 0.5cm  of c] (f1);

        \diagram* {(a) -- [] (b), 
          (a) -- [opacity=0] (a1),
          (b) -- [scalar, half right, looseness=0.7] (e), 
          (b) -- [scalar, half left, looseness=0.7] (e), 
          (b) -- [scalar, half right, looseness=0.7] (e1), 
          (b) -- [scalar, half left, looseness=0.7] (e1), 
           (b) -- [scalar, half left, looseness=0](c), 
          (c) -- [scalar, half right, looseness=0.7] (f), 
          (c) -- [scalar, half left, looseness=0.7] (f),
          (c) -- [scalar, half right, looseness=0.7] (f1), 
          (c) -- [scalar, half left, looseness=0.7] (f1), 
          (c) -- [] (d),
          };
    \end{feynman}
    \end{tikzpicture}
  \end{minipage}
  \centering
  \begin{minipage}[b]{0.4\textwidth}
   \begin{tikzpicture}[scale=3, transform shape]
   \begin{feynman}
        \vertex (a); 
        \vertex [right = 0.4cm  of a] (b);
        \vertex [right = 1.3cm  of b] (c);
        \vertex [right = 0.4cm  of c] (d);

        \diagram* {
          (a) -- [] (b), 
          (b) -- [scalar, half right, looseness=0.4] (c), 
          (b) -- [scalar, half left, looseness=0.4](c),
          (b) -- [scalar, half right, looseness=1] (c), 
          (b) -- [scalar, half left, looseness=1](c),
          (b) -- [scalar, half left, looseness=0](c),
          (c) -- [] (d)
          };
   \end{feynman}
   \end{tikzpicture}
  \end{minipage}
  \caption{The 3 different 4--loop terms for $n=5$.}\label{fig:n5} 
\end{figure}

For $\mathcal{R}=\mathcal{R}_G^5$, there are three diagrams, each at 4-loop order as shown in \cref{fig:n5}. The dressed vertex diagram (with 2 loops dressing each vertex) has the simple form
\begin{equation}
    P_{\rm DV}=\frac{(5!)^2}{(2^2\times2)^2} \prod\limits_{i=1}^{4}\left( \frac{d^3q_i}{(2\pi)^3} P_G(q_i)\right)P_G(k),
\end{equation}
which we can rewrite as
\begin{equation}
{\mathcal P}_{\rm mixed}=  225 A^5_G k_* \delta (k-k_*)
\end{equation}
and hence this term has variance 225 $A^5_G$.

The intermediate term, which corresponds to a mixed diagram with each vertex dressed by 1 loop combined with the 2-loop convolution integral as solved in this paper for the Gaussian cubed case corresponds to
\begin{equation}
    P_{\rm mixed}=\frac{(5!)^2}{2^2\times3!} \prod\limits_{i=1}^{2}\left( \frac{d^3q_i}{(2\pi)^3} P_G(q_i)\right)\int \frac{d^3q_3}{(2\pi)^3}\frac{d^3q_4}{(2\pi)^3} P_G(q_3)P_G(q_4)P_G(|\vec{k}-\vec{q_3}-\vec{q_4}|),
\end{equation}
which we can rewrite as 
\begin{equation}
    \mathcal{P}_{\rm mixed}=300 A_G^5 p(k) ,
\end{equation}
where $p(k)$ was defined by \cref{eq:p-definition} and which has variance 2, so the complete variance of the mixed term is 600.
Hence the final diagram, which has the sunset form without dressed vertices and whose form is given by \cref{eq:sunset-general} must have variance $(945-600-225)A_G^5=120A_G^5=5! A_G^5$.

\printbibliography
\end{document}